\documentclass[prb,aps,twocolumn,showpacs]{revtex4}

\usepackage{amssymb}
\usepackage{amsmath}
\usepackage{amsfonts}
\usepackage[dvips]{graphicx}
\usepackage{graphics}

\begin{document}

\bibliographystyle{apsrev}

\newcommand {\abs}[1]{| #1 |}
\newcommand {\bra}[1]{\langle \: #1 \: |}
\newcommand {\ket}[1]{| \: #1 \: \rangle}
\newcommand {\dotp}[3]{\langle \: #1 | \: #2 \: | \: #3 \: \rangle}
\newcommand {\expect}[1]{\langle \: #1 \: \rangle}
\newcommand {\ybco}[1]{YBa$_2$Cu$_3$O$_{#1}$}
\newcommand {\lacuo}{La$_2$CuO$_4$}
\newcommand {\cuo}{CuO$_2$}
\newcommand {\ybc}{YBa$_2$Cu$_4$O$_8$}
\newcommand {\Tone}{$^{63}$T$_1$}
\newcommand {\Tonec}{$^{63}$T$_{1c}^{-1}$}

\title{Charge and spin density distributions around Zn impurities in cuprates}

\author{C. Bersier, S. Renold, E. P. Stoll, and P. F. Meier}

\affiliation{Physics Institute, University of Zurich, CH-8057 Zurich, 
             Switzerland}
\date{\today}
\begin{abstract}

The effect of zinc substitution on the local electronic structure of several cuprates is investigated using first-principles cluster calculations. Clusters comprising 5, 9, and 13 copper atoms in the cuprate plane of La$_2$CuO$_4$, YBa$_2$Cu$_3$O$_7$, and YBa$_2$Cu$_4$O$_8$ are used. Spin polarized calculations with different multiplicities in the framework of density functional theory enable a detailed study of the changes in the charge and spin density distribution induced by Zn substitution. Furthermore, doping with charge carriers in the above materials is simulated and the resulting changes in the charge distribution are compared to the changes induced by Zn impurities. These differences are then discussed in terms of a phenomenological model related to properties expected from the generic phase diagram. The effects of zinc substitution are rather local and as expected the absolute values of the Mulliken charges at both nearest and next nearest neighbor oxygens to Zn are larger than in the unsubstituted clusters. The calculated electric field gradient at Cu sites that are nearest neighbor to Zn is found to be somewhat larger than in the unsubstituted cluster whereas that of next nearest neighbors is about 5~\% smaller.
We conclude that the satellite peak in the Cu NQR spectrum occurring upon Zn substitution in YBa$_2$Cu$_3$O$_7$ and YBa$_2$Cu$_4$O$_8$ has its origin at Cu that are next nearest neighbors to Zn.

\end{abstract}
 
\maketitle


\section{Introduction}

It is well known that the addition of a small amount of magnetic impurities destroys superconductivity in conventional superconductors but substitution of non-magnetic impurities is in general harmless. In hole doped cuprate superconductors, however, non-magnetic impurities like Zn and others are detrimental~\cite{bib:xiao1990} and superconductivity disappears with a few percent of Zn impurities. This fact has initiated an enormous activity in experimental and theoretical investigations on the effects of impurities that substitute copper in the CuO$_2$ planes. In spite of the great experimental and theoretical effort which has been done so far, the mechanism of the destruction of superconductivity by Zn impurities in the cuprates is still controversial.

In contrast to the vast amount of theoretical investigations on the change in all possible {\it collective} phenomena that may occur upon Zn substitution, only few informations are available about the changes in the {\it local} electronic structure.
Large supercells would have to be used in traditional band structure calculations to cope with the problems of broken lattice periodicity.
Since already in the unsubstituted materials, the unit cells contain a large number of atoms, such supercell calculations are prohibitively complex.

In the present work, the electronic structure around a Zn atom substituting a Cu is studied by ab-initio cluster calculations. These methods which are not based on periodic lattice structures, are particularly suited to calculate the changes in the charge and spin density distributions that are occurring upon the replacement of a regular atom by a new species.
Cluster calculations for Zn and Ni substituted YBa$_2$Cu$_3$O$_7$ have previously been reported by Kaplan {\it et al.}~\cite{bib:Kaplan} using clusters with five copper atoms in the CuO$_2$ plane for the pure material. They employed M\o ller-Plesset perturbation theory to investigate on the differences in the atomic charges between unsubstituted and impurity substituted clusters.
Here we present results obtained on larger clusters comprising 5, 9, and 13 copper atoms in the CuO$_2$ plane. Spin-polarized calculations within the framework of density functional theory are carried out and their accuracy is tested by comparing theoretical values for electric field gradients (EFGs) with those derived from experiments~\cite{bib:Itoh03,bib:Williams2001}. \\
It is well known that the doping of the cuprates by holes destroys the long-range antiferromagnetic correlations. On the other hand, it has been proposed that Zn substitution enhances local antiferromagnetic correlations~\cite{bib:Julien2000,bib:Itoh03}. Abrikosov~\cite{bib:Abrikosov03} referring to the calculations of Kaplan {\it et al.}\cite{bib:Kaplan} proposed that Zn, creating an excess positive charge at the copper site, reduces the hole concentration in its vicinity. He pointed out that this would lead to the appearance of a bubble of an insulating spin density wave phase around the impurity. We therefore investigate here in detail the redistribution of charge and spin densities around Zn atoms in {\lacuo }, {\ybco{7}}, and {\ybc}. In Sec.~\ref{sec:cluster} the cluster method will be described. In Secs.~\ref{sec:spindensity} and~\ref{sec:chargedensity} the resulting changes in the spin and charge density distribution induced by Zn are given. 
In addition, in Sec.~\ref{sec:chargedensity} simulations of doping with charge carriers are performed and the changes in the charge distribution are compared to the changes induced by Zn substitution. These differences are then discussed in terms of a phenomenological model related to properties expected from the generic phase diagram.
Sec.~\ref{sec:siteassignments} is devoted to the calculation of EFGs at Cu sites near Zn impurities and their comparison to experiments. Sec.~\ref{sec:summary} contains a summary and conclusions.

\section{The cluster Method and Computational Details }
\label{sec:cluster}

A cluster is a careful selection of a contiguous group of ions within a solid. The specific choice of the atoms that make up a cluster is such that it allows predominantly localized properties of a target atom and its vicinity to be calculated. A cluster consists of three regions. The target atom and at least its nearest neighboring atoms form the center of the cluster and the corresponding electrons are treated most accurately using first-principles all-electron methods. This core region is embedded in a large cloud of a few thousand point charges at the respective lattice sites imitating the Madelung potential. Point charges at the border of the core region are replaced by basis-free pseudopotentials to improve the boundary conditions for the electrons in the cluster core. These pseudopotentials make up the so-called screening region.

It would be desirable to have clusters that contain as many atoms as possible in the core region but there are two computational limitations to the cluster size: the available computer resources and the convergence of the self consistent field procedure. However, these limits have been pushed further since our first use of the cluster technique (see Ref.~\cite{bib:suter97}) which now allows to use larger clusters and hence to observe effects which could not be seen before.

\begin{table}[htb]
\begin{center}
\begin{tabular}{llccc}
\hline
Compound               &  Cluster                                      & N  & E   & B    \\ \hline
La$_2$CuO$_4$          & Cu$_{5}$O$_{26}$/Cu$_{8}$La$_{34}$            & 31 & 395 & 533  \\ \hline
La$_2$CuO$_4$          & Cu$_{9}$O$_{42}$/Cu$_{12}$La$_{50}$           & 51 & 663 & 897  \\
YBa$_2$Cu$_3$O$_{7}$   & Cu$_{9}$O$_{33}$/Cu$_{30}$Y$_{16}$Ba$_{16}$   & 42 & 573 & 780  \\
YBa$_2$Cu$_4$O$_8$     & Cu$_{9}$O$_{33}$/Cu$_{21}$Y$_{16}$Ba$_{16}$   & 42 & 573 & 780  \\ \hline

La$_2$CuO$_4$          & Cu$_{13}$O$_{62}$/Cu$_{12}$La$_{74}$          & 75 & 971 & 1313 \\
YBa$_2$Cu$_3$O$_{7}$   & Cu$_{13}$O$_{49}$/Cu$_{38}$Y$_{24}$Ba$_{24}$  & 62 & 841 & 1144 \\
YBa$_2$Cu$_4$O$_8$     & Cu$_{13}$O$_{49}$/Cu$_{25}$Y$_{24}$Ba$_{24}$  & 62 & 841 & 1144 \\ \hline
\end{tabular}
\end{center}
\caption{Compilation of the used clusters and some of their defining properties: number of atoms (N), number of electrons (E), number of basis functions (B).}
\label{tbl:compilation_II}
\end{table}

\begin{figure}[htb]
\resizebox{0.48\textwidth}{!}{
  \includegraphics{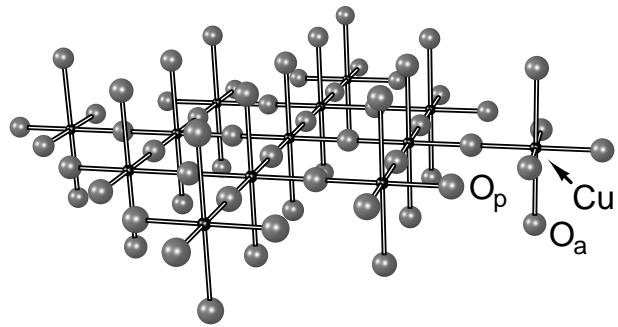}
}
\caption{Schematic representation of the core region of the Cu$_{13}$O$_{62}$/Cu$_{12}$La$_{74}$ cluster consisting of 13 copper, 36 planar, and 26 apical oxygens, all represented by basis functions.}
\label{fig:Cu13apex}
\end{figure}

\begin{figure}[htb]
\resizebox{0.48\textwidth}{!}{
  \includegraphics{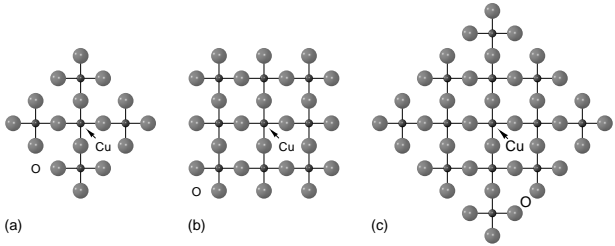}
}
\caption{Planar copper and oxygen atoms of the core region (without apical oxygens) for the clusters comprising 5(a), 9(b), and 13(c) planar copper atoms.}
\label{fig:Cu13}
\end{figure}

 In this work three clusters of different size (with 5, 9, and 13 copper atoms in the core region) are studied and their constitutive properties are given in Table~\ref{tbl:compilation_II}. For clarity these clusters are labeled X/Y where X is the chemical formula of the core region and Y the formula of the ions of the screening region represented  by pseudopotentials. Fig.~\ref{fig:Cu13apex} depicts the core region of the  Cu$_{13}$O$_{62}$/Cu$_{12}$La$_{74}$ cluster and Fig.~\ref{fig:Cu13} shows the planar atoms in the core regions of the differently sized clusters.    

The atomic positions were chosen according to crystallographic structure determinations. For {\lacuo} we used the tetragonal structure with lattice constants $a=3.77$~{\AA} and $c=13.18$~{\AA} and with a Cu-O(apex) distance of 2.40 {\AA } as given in Ref.~\cite{bib:poole1}. For {\ybco{7} } and {\ybc } the orthorhombic structure data from Ref.~\cite{bib:bordet} and Ref.~\cite{bib:fischer}, respectively, were used.
For the atoms in the core region the standard 6-311G basis sets were employed. The electronic structures in all the clusters described above were determined with the Gaussian~03 quantum chemistry package~\cite{bib:g98} in the framework of density functional theory incorporating the exchange functional proposed by Becke~\cite{bib:becke1,bib:becke2} together with the correlation functional of Lee, Yang, and Parr~\cite{bib:lyp} (specified by the  BLYP keyword in the Gaussian~03 program).
We note that band structure calculations with the local density approximation (LDA) give no magnetic solution for La$_2$CuO$_4$. In cluster calculations, however, the spin state can be chosen and in all cases (irrespective of the functional (LDA or generalized gradient approximation)) the state with the lowest variational energy is found for the antiferromagnetic arrangement (see Refs.~\cite{bib:ybcopaper,bib:martin1997}).

The results of each calculation were examined with the Mulliken population analysis which gives a description of the charge and spin densities in terms of the individual atoms and also the constituent orbitals. Other properties such as EFGs and hyperfine fields were also recorded at each copper center.

We have shown previously that with the cluster method all contributions to the EFG can be determined and the resulting values give a direct comparison to NMR and NQR experiments with good agreement~\cite{bib:ybcopaper,bib:millenniumpaper,bib:stoll}. (For planar copper atoms with vanishing asymmetry parameter $\eta$ the connection between the $z$-component of the EFG, $V^{zz}$, and the quadrupole frequency, $\nu_Q$, is given by $\nu_Q = (e/2h)QV^{zz}$, where $Q$ is the nuclear quadrupole moment.) Moreover, spin-polarized calculations enabled the evaluation of magnetic hyperfine interaction energies.

One of the great advantage of the cluster method compared with methods using periodic boundary conditions is the ability to study the effect of impurities. Band structure methods have to include large supercells to cope with the breaking of the translational symmetry. 
In contrast, the only technical complication emerging in the cluster method when introducing impurities is a possible reduction of the point group symmetry which sometimes requires more computational time.

Since we want to investigate the changes introduced by substituting a copper ion by a zinc ion we first examined the influence of lattice relaxation as the sizes of the two dications are known to be different.
Therefore, we have optimized in the  Cu$_{5}$O$_{26}$/Cu$_{8}$La$_{34}$ cluster (for {\lacuo }) the positions of the four oxygens adjacent to the central copper atom. We then repeated the same optimization procedure in a cluster where we replaced the central Cu by a Zn atom. The result is a small increase of the Zn-O distance by 0.0069~{\AA } corresponding to a tiny reduction of the EFG at the Cu adjacent to Zn of less than 0.5~\%. We therefore neglect in the following possible lattice distortions.

\section{Spin density distribution}
\label{sec:spindensity}

We have performed extensive spin-polarized calculations using all possible spin multiplicities. The general result of these quantum chemical cluster calculations is that they well reproduce the antiferromagnetic exchange interaction $J$ between NN copper ions with $\abs J$ in the range of 160 meV.

\begin{figure}
\resizebox{0.48\textwidth}{!}{
  \includegraphics{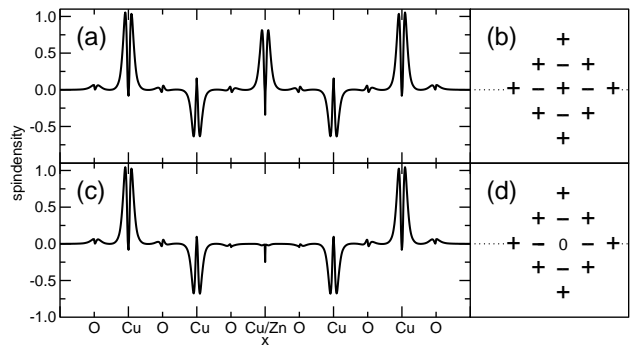}}
\caption{(a) Spin density along the 5 Cu and 6 O in the Cu$_{13}$O$_{62}$/Cu$_{12}$La$_{74}$ Cu cluster of spin multiplicity M=6; (c) spin density with Zn substitution and multiplicity M=5.}
\label{fig:spind}
\end{figure}

Here we illustrate the antiferromagnetic spin arrangement with the Cu$_{13}$O$_{62}$/Cu$_{12}$La$_{74}$ cluster for \lacuo.
With 13 copper atoms we have the possibility of choosing a spin multiplicity with an even value ranging from 2 to 14 (corresponding to a total spin $s=(M-1)/2$ ranging from $1/2$ to $13/2$). We performed all the corresponding calculations and the state with lowest energy is obtained by choosing a spin multiplicity of $M=6$.
In Fig.~\ref{fig:spind}a the spin density along the Cu-O bonds in the plane is displayed for an unsubstituted Cu$_{13}$ cluster. In the right panel (Fig.~\ref{fig:spind}b) the signs of the Mulliken spin densities at the planar Cu sites are shown which exhibit an antiferromagnetic spin arrangement, with adjacent copper sites having opposite spins. 

Most of the spin density is provided by the singly occupied molecular orbital which is also highest in energy. It is a linear combination of 3d$_{x^2-y^2}$ atomic orbitals (AOs) on the coppers ($n(3\textrm{d}_{x^2-y^2}) = 79 \%$) and 2p$_\sigma$ AOs on the oxygens ($n(2\textrm{p}_{\sigma}) = 21 \%$). The square of the 3d$_{x^2-y^2}$ AO has maxima at distances of 0.345 {\AA} from the nucleus but vanishes at the nucleus. In general, the spin density at a nuclear site is due to s AOs giving rise to the Fermi contact term. In particular, the spin densities at the Cu nuclei (cusps in Fig.~\ref{fig:spind}a) can be correlated with the Mulliken spin densities (whose signs are given in Fig.~\ref{fig:spind}b) at the same and the nearest neighbored Cu nuclei. This observation can be used to split the Fermi contact term into on-site and transferred terms as has been discussed in detail in Refs.~\cite{bib:millenniumpaper,bib:ybcopaper}. We just note here that the on-site term is negative if the Mulliken spin density at the Cu atom under consideration is positive and vice versa and that the transferred term is positive (negative) if the Mulliken spin densities at the neighboring Cu atoms is positive (negative). Near the oxygens, the spin density is provided by the 2p$_\sigma$ AO and the contact density is very small for antiferromagnetically arranged copper neighbors.  

The substitution of Zn at the cluster center introduces an extra electron into the cluster and it is to be noted that the spin state of lowest energy is now the $M=5$ state. The corresponding spin density along the Cu-O bonds (see Fig.~\ref{fig:spind}c) shows that the antiferromagnetic arrangement virtually remains intact. As expected the Mulliken spin density on the Zn is very close to zero (Fig.~\ref{fig:spind}d). Note that now the contact densities at the copper nuclei adjacent to Zn are reduced in comparison to the same sites in Fig.~\ref{fig:spind}a because there are transferred hyperfine fields from three ions only instead of four. These changes in the hyperfine fields will be further discussed in Sec.~\ref{sec:Hyperfine}.

\section{Charge density distribution}
\label{sec:chargedensity}

The cluster method enables a detailed analysis of the local charge distributions and their changes upon Zn substitution as well as upon simulations of doping. We first present the differences observed between the Cu$_{12}$ZnO$_{62}$/Cu$_{12}$La$_{74}$ and the Cu$_{13}$O$_{62}$/Cu$_{12}$La$_{74}$ clusters for La$_2$CuO$_4$. The most prominent modifications concern the oxygen atoms adjacent to Zn. We therefore then concentrate on the Mulliken charges of these oxygens and their differences in undoped and hole doped La$_2$CuO$_4$ and in optimally doped and underdoped YBa$_2$Cu$_3$O$_7$. These quantitative results are then further discussed in terms of a qualitative model of extended charges.
Finally we discuss the implications of our results for models which relate the suppression of $T_c$ upon Zn substitution with enhanced antiferromagnetism around Zn substituted regions.

\begin{figure}
\resizebox{0.48\textwidth}{!}{
  \includegraphics{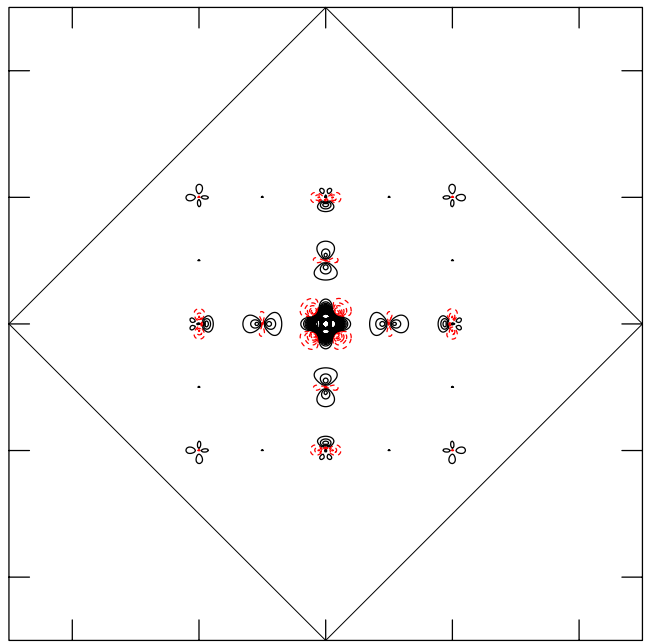}
}
\caption{(color online) Electron density difference contour plots in the cuprate plane between the Cu$_{12}$ZnO$_{62}$/Cu$_{12}$La$_{74}$ and the Cu$_{13}$O$_{62}$/Cu$_{12}$La$_{74}$ clusters on a area of $5 \times 5$ lattice constants. The tick marks correspond to the planar copper position. The solid density lines (black) occur when the local electron density of the Zn substituted cluster is larger than that of the corresponding density of the unsubstituted Cu$_{13}$ cluster. Dashed lines (red), on the other hand, denote that the local electron density is larger in the unsubstituted cluster. Two neighboring lines (solid and dashed) are separated by an electron density difference of 0.01 $e$/a$_B^3$.}
\label{fig:densitydif}
\end{figure}

We first discuss the differences in the charge distributions at all atoms. Fig.~\ref{fig:densitydif} illustrates the difference in the total electron densities in the CuO$_2$ plane between the Zn substituted and the unsubstituted {\lacuo } using calculations on the corresponding Cu$_{13}$ clusters (with the central Cu atom being substituted by Zn). In both clusters, the multiplicity leading to the lowest ground state energy was chosen: $M=6$ in the unsubstituted and $M=5$ in the Zn substituted cluster. The framed area insinuates the core region of the Cu$_{13}$ cluster. The solid (black) contours are positive electron density differences, which are more prominent than the dashed (red) contours of negative electron density differences since the central Zn atom has one electron more than the Cu atom it replaces.  The white space in the corners of the framed area is significant since the four coppers and 24 oxygens in this region are virtually unaffected by the zinc substitution and do not show up on the contour map ($\left| \Delta \rho \right| < 0.005~\textrm{a}_B ^{-3}$).

The difference between the center atoms can be explained as follows. 
As expected the 3d$_{x^2-y^2}$ AO of the Zn is almost fully occupied which shows up in the pronounced black density difference in the center of Fig.~\ref{fig:densitydif}. The red contour lines are concentrated on 3d$_{xy}$ orbitals. This is due to the fact that the 3d$_{xy}$-electrons of the Zn ion are closer to the nuclei than the ones of the Cu ions since for Zn $\left<1/r^3\right>$ = 9.39~a$_B^{-3}$ but $\left<1/r^3\right>$ = 7.94~a$_B^{-3}$ for Cu. Therefore, the electron densities at the border of the Zn ion are smaller than at the border of the Cu ion which results in red equidensity lines.

The central zinc atom of course shows the very large positive electron density difference but some has been transferred to the adjacent oxygen atoms and also to one side of the nearest neighbor copper atoms. Thus there are significantly more electrons in the 2p$_{\sigma}$ AO of the oxygens adjacent to Zn than in the unsubstituted case. The occupancy of the 2p$_{\sigma}$ orbital changes from 1.67 to 1.74 and the corresponding Mulliken charges are reduced from $-1.63$ to $-1.71$.

For the NN copper atoms we observe an asymmetric charge distribution.
The total charge on these Cu, however, almost equals that in the unsubstituted compound. On the four coppers which are next nearest neighbors to the Zn, there are more electrons in the 3d$_{x^2-y^2}$ AO compared to the bulk values.
These charge differences on the NN and NNN copper atoms imply changes in the electric field gradients which will be discussed in detail in Sec.~\ref{sec:EFG}.

The calculations in Ref.~\cite{bib:Kaplan} yield similar changes in the Mulliken charges at the oxygens adjacent to the Zn impurity but the changes at the adjacent \emph{copper} sites are much larger than those calculated in this work. This difference might be due to the use of bare point charges in the immediate vicinity of the border of the cluster. It was shown in Ref.~\cite{bib:martinhay} that especially positive point charges should be replaced by pseudopotentials since they present a too strong Coulomb potential for the electrons in the core region.

It is remarkable that the effects of Zn substitution are rather local as concerns the electronic structure. This local screening, however, tells nothing about collective phenomena like scattering where it can produce a much more extended range of influence.

We have also determined the electron density differences for clusters representing the YBa$_2$Cu$_3$O$_7$ and YBa$_2$Cu$_4$O$_8$ materials. The results are very similar to those for La$_2$CuO$_4$ shown in Fig.~\ref{fig:densitydif} and discussed above. Furthermore, to simulate the effect of hole doping in La$_2$CuO$_4$ we have repeated all calculations for the corresponding clusters Cu$_{13}$O$_{62}$/Cu$_{12}$La$_{74}$ and Cu$_{12}$ZnO$_{62}$/Cu$_{12}$La$_{74}$ with one electron less. These simulations correspond to a hole doping of roughly 8 \%. In the same way, the effects of underdoping YBa$_2$Cu$_3$O$_7$ were studied by repeating all calculations with one electron more. To discuss the essential changes we concentrate on the Mulliken charges of the four oxygen atoms adjacent to Zn or, in the unsubstituted compounds, Cu.

\begin{figure}
\resizebox{0.48\textwidth}{!}{
  \includegraphics{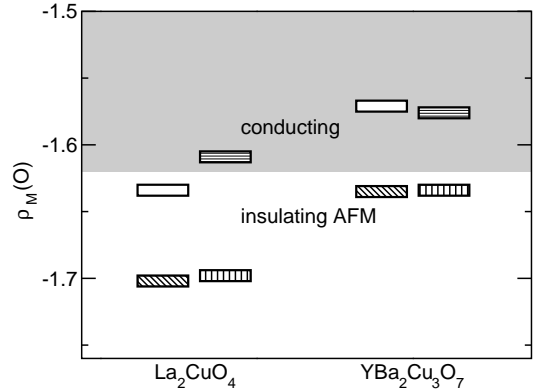}
}
\caption{Schematic view of the Mulliken charge on the oxygens in the unsubstituted compounds (white rectangles) and in Zn substituted compounds adjacent to a Zn atom (diagonally hatched rectangles) for La$_2$CuO$_4$ and YBa$_2$Cu$_3$O$_7$. The results for clusters doped with one hole for La$_2$CuO$_4$ and one electron for YBa$_2$Cu$_3$O$_7$ are plotted with horizontally (in unsubstituted clusters) and vertically hatched (in Zn substituted clusters) rectangles. An approximate border line at $-1.62$ is drawn to illustrate the separation between the insulating antiferromagnetic phase and the conducting phase.}
\label{fig:ChardensLaY}
\end{figure}

The calculated Mulliken charges of the planar oxygens in {\lacuo } and {\ybco{7} } are  $-1.64$  and $-1.57$, respectively, as illustrated by the white rectangles in Fig.~\ref{fig:ChardensLaY}. Both these values are reduced to $-1.71$ and  $-1.64$, respectively (diagonally hatched rectangles), for oxygens adjacent to substituted Zn atoms.
The simulation of hole doping in {\lacuo } results in a Mulliken charge of $-1.61$ without Zn (horizontally hatched rectangles) and $-1.70$ (vertically hatched rectangles) on the oxygens adjacent to Zn. An analogous simulation but with electron doping was performed for {\ybco{7}}. If an electron is added to the Cu$_{13}$O$_{49}$/Cu$_{38}$Y$_{24}$Ba$_{24}$ cluster the Mulliken charges on the oxygens reduces slightly to $-1.58$ in the cluster without Zn and is virtually unaffected $(-1.64)$ in the Zn substituted cluster.

These quantitative results for the {\it local} changes near substituted Zn and for the effects of simulating about 8~\% hole doping are now put in perspective with the {\it extended} changes expected from the generic phase diagram. The following discussion is therefore necessarily qualitative and completely phenomenological since we do not consider band structure. It serves, however, to rationalize the main changes that are locally produced by Zn substitution.

The conventional wisdom in cuprate superconductors is that the pure parent compounds {\lacuo } and {YBa$_2$Cu$_3$O$_6$} are insulators and that upon ``hole doping'' the missing electrons come from the planar oxygens. The holes thus introduced on the oxygens destroy rapidly the long range antiferromagnetic order and the materials become conducting. Although we have pointed out elsewhere~\cite{bib:holes,bib:holesII} that this picture is too naive and may easily lead to wrong conclusions, a semiquantitative discussion on the values of the oxygen charges and their changes with hole and electron doping both in the unsubstituted and in the Zn substituted compound is instructive.
We have therefore indicated in Fig.~\ref{fig:ChardensLaY} a schematic separation into a metallic region (gray) and an insulating region (white) with a border line at the arbitrary chosen value of  $-1.62$. 

This puts the Mulliken charge on the oxygen in undoped La$_2$CuO$_4$ (white rectangle in Fig.~\ref{fig:ChardensLaY}) into the insulating region and that obtained with hole doping in the more metallic region. From simulations of underdoped YBa$_2$Cu$_3$O$_7$, however, we find a small decrease of the Mulliken charge which still is the conducting zone.

In each of the four cases considered, Zn substitution drives the Mulliken charges at the oxygens deeper into the insulating region. In particular, Zn substitution in YBa$_2$Cu$_3$O$_7$ reduces $\rho_{M}(\textrm O)$ to a value (diagonally hatched rectangles) which is almost equal to $\rho_{M}(\textrm O)$ in undoped and Zn free La$_2$CuO$_4$ (white rectangle). Hence, we observe that the electronic structure in the immediate vicinity of a Zn impurity in hole doped cuprates is similar to that found in the Zn free and undoped parent compounds.

This schematic description of the effect of Zn substitution and charge carrier doping can be interpreted in terms of the generic phase diagram: Zn substitution overcompensates the hole doping effects in its vicinity and thereby creates an environment where the antiferromagnetic correlations are restored. The prediction, however, whether this leads to a static or dynamic spin ordering cannot be given with our cluster method.

Several experiments in Zn substituted cuprates have also been interpreted as indications for locally enhanced antiferromagnetic correlations (see Ref.~\cite{bib:Julien2000,bib:Itoh03}) induced by Zn. This has been emphasized in the work of Abrikosov~\cite{bib:Abrikosov03} who considered the existence of antiferromagnetic bubbles which surround Zn, Ni, or Li impurities in the CuO$_2$ plane on the basis of the general theory of metal-insulator transition in layered cuprates.

\section{Site assignments in copper NMR and NQR spectra}
\label{sec:siteassignments}

Numerous NMR and NQR experiments on Zn substituted cuprates have been reported. In the present work we refer in particular to the measurements of Itoh {\it et al.}\cite{bib:Itoh03} and Williams {\it et al.}\cite{bib:Williams2001} who presented $^{63}$Cu quadrupole resonance frequencies and spin-lattice relaxation rates in unsubstituted and Zn substituted YBa$_2$Cu$_3$O$_7$ and YBa$_2$Cu$_4$O$_8$. It has been observed that the main copper line broadens and slightly shifts to higher frequencies when Zn is substituted and that a new Cu satellite resonance occurs at a somewhat lower frequency. The intensity of this satellite peak grows with increasing Zn concentration.

In addition, Itoh {\it et al.}\cite{bib:Itoh03} found that the Cu spin-lattice relaxation time, $^{63}T_1$, of the satellite peak is shorter than that of the main signal. This was interpreted as resulting from locally enhanced antiferromagnetic correlations in the neighborhood of Zn based on the site assignment that the NQR satellite peak comes from Cu atoms that are nearest neighbors to a substituted Zn atom.

In this chapter we propose a different site assignment. We claim that the satellite peak is due to copper atoms that are \emph{next} nearest neighbors to a substituted Zn atom. We will provide evidence for this in Sec.~\ref{sec:EFG} by a direct calculation of copper EFGs in clusters representative of unsubstituted and Zn substituted La$_2$CuO$_4$, YBa$_2$Cu$_3$O$_7$, and YBa$_2$Cu$_4$O$_8$. In Sec.~\ref{sec:Hyperfine} we will consider measurements of copper spin-lattice relaxation rates  in unsubstituted and Zn substituted cuprates to give additional independent arguments that support the site assignment suggested above.
Furthermore, in Sec.~\ref{sec:sensitivity} a detailed investigation about the sensitivity of the EFGs on the charge distribution is given.

\subsection{Electric Field Gradients}
\label{sec:EFG}

The extra calculations needed to study the changes of the copper EFGs in substances doped with impurities have been performed -- for efficiency reasons -- in clusters comprising nine copper atoms in the plane. As already stated in Sec.~\ref{sec:chargedensity} this is justified since the changes in the electronic structure upon Zn substitution occur predominantly in the immediate vicinity of the impurity (see also Fig.~\ref{fig:densitydif}).

\begin{figure}
\begin{center}
\includegraphics[width=0.48 \textwidth]{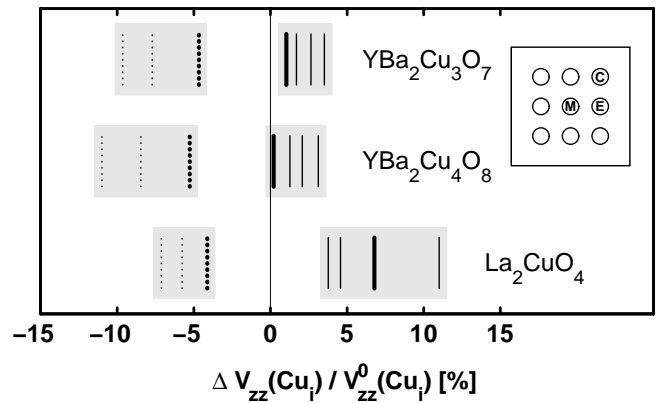}
\end{center}
\caption{Relative differences between the copper EFGs in the Zn substituted and in the unsubstituted clusters with solid (dotted) lines referring to copper atoms which are NN (NNN) to the Zn impurity.
In the inset: schematic representation of the Zn sites in the Cu$_9$ cluster with the site labelings used in the text (M: middle, E: edge, C: corner).
Thick bars are for situations where the Zn is replaced at site M whereas thin bars denote results from clusters where the impurity has been inserted at sites E or C (see text).}
\label{fig:efg_vs_nnn}
\end{figure}

We have evaluated the electronic structure in clusters for La$_2$CuO$_4$, YBa$_2$Cu$_3$O$_7$, and YBa$_2$Cu$_4$O$_8$ and recorded the copper EFGs at all copper sites in the cluster which slightly differ due to finite size effects. Next we repeated the above calculations for Zn substituted clusters where the copper in the middle of the cluster (site M in the inset of Fig.~\ref{fig:efg_vs_nnn}) was replaced by a zinc. The copper EFGs were again calculated at all sites and were compared to those at the same sites in the unsubstituted clusters.

The relative differences between the copper EFGs in the Zn substituted and in the unsubstituted clusters are shown with thick bars in Fig.~\ref{fig:efg_vs_nnn}. The solid (dotted) thick lines refer to situations where the copper in the substituted cluster is a nearest neighbor (NN) (next nearest neighbor (NNN)) to the Zn impurity.
The EFG values for the NN coppers are all found to be larger than in the unsubstituted compounds whereas those for the NNN are smaller.

For consistency, we repeated the above calculations but with Zn impurities in the edge and corner position (site E and site C).
We included the results of these calculations with thin bars in Fig.~\ref{fig:efg_vs_nnn} and -- as a guide to the eye -- grouped the copper EFGs resulting from coppers with the same distance (NN or NNN) to the Zn impurity by gray blocks. It is to be expected that asymmetrical substitution leads to somewhat deviating values, however, the qualitative findings of the above paragraph remain unchanged.

We therefore conclude that the observed satellite peak with lower NQR frequency originates from copper nuclei that are NNN to Zn. The NN will only broaden the main line and induce a shoulder towards higher frequency which is clearly visible in the Cu NQR spectrum~\cite{bib:Itoh03,bib:Williams2001}.

Note that in Refs.~\cite{bib:Itoh03,bib:Williams2001} it was suggested by probabilistic arguments that the satellite peaks may originate from copper sites with one Zn NN. However, this site assignment is not unambiguous, since the probability of finding a copper coordinated with one Zn NN is the same as that of the Zn being in a NNN position to the copper, at least for low Zn concentration.

\subsection{Comparison of the calculated hyperfine field with the NMR spin-lattice relaxation rate}
\label{sec:Hyperfine}

In this paragraph we will give additional independent arguments that support the above conclusion about the site assignments in $^{63}$Cu NQR spectra.

The spin-lattice relaxation rate with the applied field along the $c$-axis, $^{63}T_{1c}^{-1}$, is due to fluctuations of the hyperfine fields in the perpendicular directions, i.e. in the CuO$_2$ plane. Adopting the notation from the Mila-Rice spin Hamiltonian~\cite{bib:Mila}, there is an on-site hyperfine interaction at the copper nucleus, denoted by  $A_{\bot}$, and a transferred field $B$. For the case of fully antiferromagnetically correlated copper spins, the rate $^{63}T_{1c}^{-1}$, is proportional to
\begin{equation}
^{63}T_{1c}^{-1} \propto  (A_{\bot}-4B)^2
\label{eq:Tonec}
\end{equation}
since each of the four NN contributes one $B$.
A Zn NN, however, will hardly transfer any spin density to the Cu (see Fig.~\ref{fig:spind}) and we expect
\begin{equation}
 ^{63}T_{1c}^{-1} \propto  (A_{\bot}-3B)^2.
\label{eq:TonecZn}
\end{equation}

Typical values~\cite{bib:nandor} are $A_{\bot}=0.29~\mu$eV  and $B=0.4~\mu$eV, which would reduce $^{63}T_{1c}^{-1}$ by $1-(A_{\bot}-3B)^2/ (A_{\bot}-4B)^2=52~\%$. Note that if there were no correlations at all, this reduction would still be $1-(A_{\bot}^2+3B^2)/ (A_{\bot}^2+4B^2) = 22~\% $. 

\begin{figure}[htb]
\resizebox{0.48\textwidth}{!}{
  \includegraphics{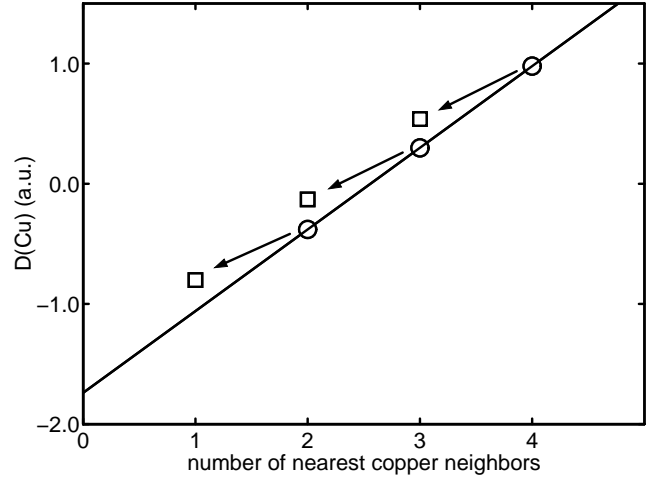}
}
\caption{Contact spin density $D(\textrm{Cu})$ in a$_B^{-3}$ vs. the number of nearest copper neighbors. The circles result from calculations for the unsubstituted compound whereas the squares refer to Zn substituted clusters.} 

\label{fig:cp_vs_nn_lacuo}
\end{figure}

To obtain a more quantitative statement we can use the results of our cluster calculations in the Cu$_9$O$_{42}$/Cu$_{12}$La$_{50}$ and Cu$_8$ZnO$_{42}$/Cu$_{12}$La$_{50}$ clusters for the {\lacuo } system. The circles in Fig.~\ref{fig:cp_vs_nn_lacuo} indicate the calculated Fermi contacts, $D(\textrm{Cu})$, in the unsubstituted Cu$_9$ cluster recorded at different positions in the cluster (see inset of Fig.~\ref{fig:efg_vs_nnn}) with different number of nearest neighbors. The slope of the fitted straight line gives the transferred contribution and the intersect the on-site contribution. The squares denote results in a Zn substituted cluster. The arrows are meant to indicate that upon losing one Cu neighbor the Fermi contact drops by about $0.6B$. Since there is some spin density on the oxygen between Zn and Cu, the change of the total transferred spin density is not from $4B$ to $3B$ but to $3.4B$ which still leads to an enormous reduction of the spin-lattice relaxation rate.

For a Cu nucleus adjacent to a Zn we therefore expect a considerably reduced spin-lattice relaxation rate. If these copper nuclei contributed to the satellite peak they would relax slower than those contributing to the main peak in contrast to the observations (see Fig. 7 in Ref.~\cite{bib:Itoh03}). In addition the spin-lattice relaxation rate of nuclei contributing to the main peak would not change when substituting Zn atoms in the cuprate plane. However, the data in Fig. 6 of Ref.~\cite{bib:Itoh03} show that, in fact, the relaxation of the copper nuclei in the main peak is slower in Zn substituted compounds. This behavior is expected only if, instead, the Cu atoms nearest neighbored to Zn contribute to the main peak.

We thus conclude that the satellite peak at which a larger rate is observed cannot be attributed to a Cu which is nearest neighbor to Zn thus providing an additional argument to the assignment made in Sec.~\ref{sec:EFG}.

\subsection{Sensitivity of EFG values on charge distribution}
\label{sec:sensitivity}

A detailed description of the evaluation of EFG in the framework of cluster calculations is found in Ref.~\cite{bib:stoll}. Here we just point out the relevant aspects that are connected to the rearrangement of charge distributions in the neighborhood of a Zn impurity.
In cluster calculations, the charge distribution is determined by the occupied MOs. The $m^{th}$ MO is represented as a linear combination

\begin{equation}
\phi_m(\vec{r})=\sum_{K=1}^n\phi_m^K(\vec{r}-\vec{R}_K)=\sum_{K=1}^n\sum_{k=1}^{n_K}c_m^{K,k}B_{K,k}(\vec{r}-\vec{R}_K)
\label{eq:linearcomb}
\end{equation}
of $n_K$ atomic basis functions $B_{K,k}$ centered at the nuclear sites $K=1,\dots,n$, and the $c_m^{K,k}$ are the MO coefficients.
We assume in the following that the target nucleus $K_T$ is at $\vec{R}_{K_T}=0$. The contribution of the MO $\phi_m$ to the EFG at $K_T$ is given by the matrix element
\begin{equation}
V_m^{ij}=\dotp{\phi_m(\vec{r})}{\frac{3x^ix^j-r^2\delta^{ij}}{r^5}}{\phi_m(\vec{r})}
\nonumber
\end{equation}
\begin{equation}
=\sum_{K=1}^{n}\sum_{L=1}^{n}\sum_{k=1}^{n_K}\sum_{l=1}^{n_L}c_m^{K,k}c_m^{L,l} \times
\label{eq:matrixel}
\nonumber
\end{equation}
\begin{equation}
\times \dotp{B_{K,k}(\vec{r}-\vec{R}_K)}{\frac{3x^ix^j-r^2\delta^{ij}}{r^5}}{B_{L,l}(\vec{r}-\vec{R}_L)}.
\end{equation}

Thus we can identify three types of contributions: $(i)$ on-site terms from basis functions centered at the target nucleus ($K=L=K_T$), $(ii)$ mixed on-site and off-site contributions, and $(iii)$ purely off-site terms with $K\neq K_T$ and $L \neq K_T$ (see Ref.~\cite{bib:stoll} for further details).

To rationalize the changes of the EFG on the copper sites upon replacing a Cu by a Zn atom, it is sufficient to concentrate on the MOs which contain partially occupied  3d$_{x^2-y^2}$ and 3d$_{3z^2-r^2}$ AOs at the target nucleus.
The EFG is then given by the contributions $R$ from all other orbitals and these two frontier orbitals:
\begin{equation}
V^{zz}=R+\frac{4}{7} ( N_{3z^2-r^2}\expect{r^{-3}}_{3z^2-r^2} - N_{x^2-y^2}\expect{r^{-3}}_{x^2-y^2})
\label{eq:EFG}
\end{equation}
Here, $N$ are the occupancies of the corresponding orbitals. In Table~\ref{tbl:occupancies} we collect the values obtained at the NNN Cu in the pure and the Zn substituted clusters, respectively. We illustrate this fact with the \lacuo\ substance and note that the contributions from other orbitals ($R$) is marginally changed upon Zn substitution.

The decrease of 4~\% of $V^{zz}$ at the NNN Cu is thus due to an increase in $N_{x^2-y^2}$ by 0.010 and a decrease in $N_{3z^2-r^2}$ by 0.002. This demonstrates that EFGs are rather sensitive to deviations of the electronic charge densities. 

The fact that the partial occupancies $N$ of the AOs are very similar to the partial Mulliken charges $\rho_M$ which are also included in Table~\ref{tbl:occupancies} gives evidence that despite their simplistic concept the Mulliken charges can be significant quantities in the analysis of charge distributions in cuprates.

\begin{table}[htb]
\begin{center}
\begin{tabular}{lcccc}
\hline
\multicolumn{1}{c}{} & \multicolumn{2}{c}{unsubstituted} & 
\multicolumn{2}{c}{Zn substituted} \\
& $3d_{x^2-y^2}$ & $3d_{3z^2-r^2}$ & $ 3d_{x^2-y^2}$ & $3d_{3z^2-r^2}$ \\ \hline
$N$                       & 1.334 & 1.921 & 1.344 & 1.919 \\
$\rho_M$                  & 1.377 & 1.931 & 1.387 & 1.929 \\
$\expect{r^{-3}}$         & 8.058 & 8.004 & 8.048 & 7.999 \\
\hline\end{tabular}
\caption{Occupancies $N$, partial Mulliken charges $\rho_M$, and expectation values $\expect{r^{-3}}$ for the relevant copper AOs in unsubstituted ($M=2$) and Zn substituted ($M=1$) Cu$_9$ clusters representative of La$_2$CuO$_4$.}
\label{tbl:occupancies}
\end{center}
\label{tbl:MullikenchargediffYBCO}
\end{table}

\section{Summary and Conclusions}
 
\label{sec:summary}

The objective of this work was a first-principles investigation on the effect of Zn impurities that substitute Cu atoms in the CuO$_2$ plane of the cuprates on the local electronic structure. This was performed in the framework of spin-polarized density functional theory with localized basis functions in clusters representative of La$_2$CuO$_4$, YBa$_2$Cu$_3$O$_7$, and YBa$_2$Cu$_4$O$_8$ and the respective Zn substituted compounds. As a significant improvement of our previous work~\cite{bib:millenniumpaper,bib:ybcopaper} we were able to include up to 13 planar copper atoms together with the surrounding planar and apical oxygen atoms in our clusters.

For the unsubstituted compounds the antiferromagnetic spin arrangement with the Mulliken spin densities of neighboring copper atoms having opposite sign could well be reproduced. Calculations in clusters where a Cu was replaced by a Zn atom revealed almost no spin density at the Zn site whereas the antiferromagnetic spin alignment around the impurity was left virtually intact.

Using the same clusters, we observed in general that the changes in the charge distribution upon Zn substitution remain rather local and does not extend to more than the next nearest copper atoms. In particular the Mulliken charge at the oxygens adjacent to the Zn impurity was significantly reduced to more negative values compared to the Zn free case. We thus obtain a reduced ``hole concentration'' in the immediate neighborhood of a Zn impurity, i.e. on the four next nearest oxygen atoms, i.e. on the four next nearest oxygen atoms. This result is in agreement with that obtained previously by Kaplan {\it et al.}~\cite{bib:Kaplan}.
For a comparison of the effects of Zn substitution with those of hole doping, we then simulated a hole doping of about 8~\% in unsubstituted La$_2$CuO$_4$ by removing one electron from the Cu$_{13}$ cluster and observed that in general impurity substitution seems to compensate the effects of hole doping on the charge distribution. We interpreted this as evidence for enhanced antiferromagnetic correlations around Zn impurities.

Furthermore, we also calculated electric field gradients at Cu sites in unsubstituted and Zn substituted clusters and found that the EFGs at Cu sites which are nearest neighbors to a Zn impurity are slightly larger than in the unsubstituted compounds whereas those at sites NNN to Zn are about 5~\% smaller. We therefore proposed a new site assignment for Cu NQR spectra in Zn substituted compounds: the satellite peak at lower frequencies originates from NNN Cu to Zn whereas the NN Cu induce a shoulder in the main peak towards higher frequencies. This site assignment is corroborated by arguments concerning the different spin-lattice relaxation rates of nuclei in the main and the satellite peak.

\begin{acknowledgments}
Special thanks go to M. Mali, J. Roos, and T. Feh\'er for numerous stimulating discussions. Parts of this work were done in collaboration with T. A. Claxton whose help and critical discussions are acknowledged. This work was partially supported by the Swiss National Science Foundation.
\end{acknowledgments}


\end{document}